\documentclass[12pt]{article}
\usepackage{graphicx}
\usepackage{amsmath,amssymb}
\usepackage{float}
\usepackage{cite}

\newcommand{\p}{\bot}
\newcommand{\dd}{\partial}
\newcommand{\de}{\delta}
\newcommand{\om}{\omega}

\newcommand{\ls}{\left(}
\newcommand{\rs}{\right)}
\newcommand{\ti}{\tilde}
\newcommand{\g}{\gamma}
\newcommand{\ff}{\varphi}
\newcommand{\la}{\lambda}
\newcommand{\m}{\mu}
\newcommand{\ee}{\eta}
\newcommand{\disn}[2]{$$\displaylines{\refstepcounter{equation}\label{#1}\hskip
1em minus 1em #2\hfilneg}$$}
\newcommand{\nom}{\hfil\hskip 1em minus 1em (\theequation)}
\newcommand{\ns}{\hfill\cr\hfill}
\newcommand{\no}{\hfil \hskip 1em minus 1em\phantom{(\theequation)}
            \hfilneg\cr\hfilneg\hskip 1em minus 1em\hfil}

\usepackage{caption2}

\begin{document}

\title{Renormalized Light Front Hamiltonian\\ in the Pauli-Villars
Regularization}

\author{
M.Yu.~Malyshev\thanks{E-mail: mimalysh@yandex.ru},
S.A.~Paston\thanks{E-mail: paston@pobox.spbu.ru},
E.V.~Prokhvatilov\thanks{E-mail: evgeni.prokhvat@pobox.spbu.ru},
R.A.~Zubov\thanks{E-mail: roman.zubov@hep.phys.spbu.ru}\\
{\it Saint Petersburg State University, St.-Petersburg, Russia}
}
\date{\vskip 15mm}
\maketitle

\begin{abstract}
We address the problem of nonperturbative calculations on the light
front in quantum field theory regularized by Pauli-Villars method.
As a preliminary step we construct light front Hamiltonians in
(2+1)-dimensional $\lambda \varphi^4$ model, for
the cases without and with spontaneous symmetry breaking. The
renormalization of these Hamiltonians in Pauli-Villars regularization
is carried out via comparison of all-order perturbation theory,
generated by these Hamiltonians, and the corresponding covariant
perturbation theory in Lorentz coordinates.
\end{abstract}

\newpage

\section{Introduction}
\label{intro}
Hamiltonian formulation on the light front (LF) \cite{dir} leads in quantum field theory
to simple description of the vacuum state, that simplifies the
nonperturbative Hamiltonian approach to the bound state and mass spectrum problem \cite{paul,
Bakker.Bassetto.Brodsky.Broniowski.Dalley.Frederico.Glazek.
Hiller.Ji.Karmanov.et.al.arXiv2013}.
The LF can be defined by the equation $x^+=0$ where
$x^{+} = \frac{x^0 + x^1}{\sqrt{2}}$ plays the role of time ($x^0, x^1, x^{\p}$ are Lorentz
coordinates with
 $x^{\p}$ denoting the remaining spatial coordinates). The role of usual space coordinates is
 played
by the LF coordinates $x^{-} = \frac{x^0 - x^1}{\sqrt{2}}, \, x^{\p}$.

The generator $P_-$ of translations in $x^-$ is kinematical \cite{dir} (i.e. it is independent of the
interaction and quadratic in fields, as a momentum in a free theory). It is nonnegative ($P_-\geqslant0$)
for quantum states with nonnegative mass squared. So the state with the minimal eigenvalue $p_-=0$ of the
momentum operator $P_-$ can describe (in the case of the absence of the massless particles) the vacuum
state, and it is
also the state minimizing the $P_+$ in Lorentz invariant theory. Furthermore it is possible
to introduce the Fock space on this vacuum and  formulate in this space the eigenvalue problem for the
operator $P_+$
(which is the LF Hamiltonian) and find the  spectrum of mass $m$ in subspaces with fixed values of the
momenta $p_-, p_{\bot}$ \cite{paul, NPPF}\footnote{Review \cite{NPPF} includes the necessary (for present
work) content of papers \cite{Ilgen, tmf97, tmf99, tmf02}.}:
\disn{1aa}{
P_+|p_-,p_{\bot}\rangle=\frac{m^2+p_{\bot}^2}{2p_-}|p_-,p_{\bot}\rangle.
\nom}

The theory on the LF has the singularity at $p_- = 0$,
and the simplest regularization is the cutoff $p_-\geqslant\de>0$.
Other convenient translationally invariant regularization, that can treat also
 zero ($p_-=0$) modes of fields, is the cutoff $|x^-| \leqslant L$ plus
periodic boundary conditions for fields. This regularization
discretizes the momentum $p_-$ ($p_-=\frac{\pi n}{L} ,\,
n=0,1,2,...$) and clearly separates zero and nonzero modes. It is
the so-called "Discretized Light Cone Quantization" (DLCQ). Such
regularization was successfully used to solve the problem
(\ref{1aa}) for (1+1) field theories: Sine-Gordon model
\cite{Annenkova}, Yukawa model \cite{Brodsky.Pauli}, Quantum
Electrodynamics (QED) \cite{Eller.Pauli.Brodsky} and Quantum
Chromodynamics (QCD) \cite{Hornbostel.Brodsky.Pauli}. The
significant  perturbative analysis of LF (or infinite-momentum
frame) gauge theory in (3+1)-dimensions was made in papers
\cite{Brodsky.Roskies.Suaya} and \cite{Srivastava.Brodsky2001}.
Nevertheless the problem  of constructing the renormalized LF
Hamiltonians using, in particular, the above-mentioned
regularizations turned out to be very difficult. We refer to
nonperturbative Similarity Renormalization Group (SRG) approach
\cite{Wilson, Glazek, Glazek.Wilson.Phys.Rev.D1993,
Glazek.Wilson.Phys.Rev.D1994, Glazek.ActaPhys.Polon.B2008} which
allows to construct approximately effective LF Hamiltonians acting
in the space of small number of effective (constituent) particles
\cite{Glazek.Phys.Rev.D2001, Glazek.Wieckowski.Phys.Rev.D2002}.

All used regularizations of the singularity at $p_- = 0$ are not Lorentz invariant.
This can lead to nonequivalence of the results
obtained with
the LF and the conventional  formulation in Lorentz coordinates.
It was shown in papers \cite{burlang, tmf97,NPPF} that some diagrams
of the perturbation theory, generated by the LF Hamiltonian, and
corresponding diagrams of the conventional  perturbation theory in
Lorentz coordinates can differ.
In  papers \cite{tmf97, NPPF} it was found how to restore the equivalence of the LF and
conventional perturbation theories in all orders in the coupling constant by addition of new
(in particular, nonlocal) terms to the canonical LF Hamiltonian. These terms must remove the
above-mentioned
differences of diagrams.

The method of the restoration of the equivalence between the LF and
conventional perturbation theories, found in \cite{tmf97, NPPF}, was applied to constructing of correct
renormalized LF Hamiltonian for (3+1)-dimensional Quantum Chromodynamics \cite{tmf99, NPPF}.
In the papers \cite{tmf02, NPPF} this method was applied to massive Schwinger model
((1+1)-dimensional
Quantum Electrodynamics) and correct LF Hamiltonian was constructed. This Hamiltonian was used for
numerical calculations
of the mass spectrum \cite{Yad.Fiz.2005}, and the obtained results well agree with lattice
calculations in Lorentz coordinates \cite{Hamer2000}
for all values of the coupling (including very large ones).

The number of the above-mentioned new terms, which must be added to canonical LF Hamiltonian, and
counterterms,
necessary for the ultraviolet (UV) renormalization,
depends essentially on the  regularization scheme. For the case of QCD(3+1)
 \cite{tmf99, NPPF}  in the light-cone
gauge one gets the finite number of these terms only in the regularization of the Pauli-Villars (PV)
 type \cite{PV}. This regularization violates gauge invariance. However it was shown in \cite{tmf99,
NPPF}
that  gauge invariance can be restored in  renormalized  LF theory with proper choice of coefficients
before these new terms and counterterms.
 On the other side, the PV regularization involves the introduction of auxiliary ghost fields (with
the large  mass playing the
role of the regularization parameter). These ghost fields generate the states with
the indefinite metric, and one has to deal with such states in
the nonperturbative (e.g. variational) calculations using the LF Hamiltonian.
Attempts to do these calculations were made in papers
\cite{Brodsky1, Brodsky2, Brodsky3, Brodsky4} for nongauge theories.
It is important to generalize this for gauge theories like QCD (e.g. for the formulation
 \cite{tmf99, NPPF}, where the PV regularization introduces ghost gauge fields).

Calculations with truncated LF Fock basis within PV regularization
were carried out in
\cite{Brodsky.Franke.Hiller.McCartor.Paston.Prokhvatilov2004,
Hiller.Chabysheva.Phys.Rev.D2010,
Hiller.Chabysheva.Phys.Rev.D82.2010,
Hiller.Chabysheva.Phys.Rev.D2011}. The generalization of this
method allowing to consider the states with infinite number of
quanta was proposed in \cite{Hiller.Chabysheva.Phys.Lett.B2012,
Hiller.Chabysheva.arXiv2012} (it is the so called LF
coupled-cluster (LFCC) method). The PV regularization was used, in
particular,  in Covariant Light Front Dynamics (CLFD) approach
\cite{Carbonell.Desplanques.Karmanov.Mathiot.PhysRep1998,
Mathiot.Smirnov.Tsirova.Karmanov.FewBodySyst2011,
Karmanov.Mathiot.Smirnov.PhysRevD2010} where the truncation of LF Fock basis of states and
the corresponding renormalization procedure within covariant formulation
on the LF were used.

The question of using the PV regularization in the
LF Hamiltonian approach isn't studied sufficiently.
So we address this question in the present paper. For the investigation of the problem we start
 with the construction of
the renormalized LF Hamiltonian in the PV regularization for the scalar field theory in
the (2+1)-dimensional space-time.

We compare the perturbation theory generated by the LF Hamiltonian and covariant perturbation
theory in Lorentz
coordinates by the method of papers \cite{tmf97, NPPF}. This allows to find the counterterm
necessary for the renormalization of the LF Hamiltonian by the calculation of the divergent
part of the corresponding diagram in the covariant perturbation theory in Lorentz coordinates.
Let us note that there is the possibility to carry out the renormalization directly in $x^+$-ordered
perturbation theory \cite{Brodsky.Roskies.Suaya}. However the renormalized theory on the LF at that
approach can, in principle, turn out to be nonequivalent to the
 original Lorentz covariant theory due to possible differences between  finite diagrams generated by the
LF Hamiltonian and corresponding to them covariant diagrams.

To take into account the different vacua appearing in considered model due to the spontaneous
symmetry breaking we consider the transition to the LF Hamiltonian from the theories quantized on
the spacelike
planes approaching to the LF. In these theories it is possible to determine the true vacuum
using the Gaussian approximation \cite{Stev}. Accordingly we get two different expressions for the
LF Hamiltonian for the cases without and with the spontaneous symmetry breaking.
Let us note that this problem can be related to zero mode problem  on the LF
\cite{Hornbostel.Phys.Rev.D1992}. Also we note that the description of spontaneous symmetry breaking
within the theory
on the LF was considered for the Standard Model in the paper \cite{Srivastava.Brodsky2002}.

In Sect.~\ref{sec:1} we formulate the scalar field theory in the coordinates corresponding to the
spacelike planes approaching to the LF and solve the vacuum problem in the Gaussian
approximation. In Sect.~\ref{sec:2} we investigate the perturbation theory for this model. Using
the PV regularization we prove
the coincidence of the diagrams of the perturbation theory, generated by the LF Hamiltonian,
and the corresponding diagrams of the covariant perturbation theory in Lorentz coordinates
in the limit of removing the LF momentum cutoff (i.e. $\de\to0$).
In the concluding Sect.~\ref{concl} we consider a way to solve the eigenvalue problem for obtained LF Hamiltonians. Also we discuss the possible generalization of this way to QCD and the approach related to the AdS/QCD correspondence
\cite{Brodsky.Teramond.Phys.Rev.Lett.2006, Brodsky.Teramond.Phys.Rev.Lett.2009}.
Appendix A contains
the calculation of the divergent part
of the diagram that defines the counterterm which renormalizes the theory. Appendix B gives the
example of a comparison of diagram calculations in LF
and conventional covariant formulations.

\section{Light front Hamiltonian construction for
the scalar field theory in (2+1)-dimensional space-time}
\label{sec:1}
To clarify the way of the construction of the LF Hamiltonian we start from the Lagrangian
formulation in the coordinates $y^{\m}$
approaching the LF coordinates $x^{\m} = (x^+, x^-, x^{\p}):$
\disn{3}{
 y^0 = x^+ + \frac{\,\ee^2}{2} \, x^-, \,\, y^1 = x^-, \,\, y^{\p} = x^{\p},
\nom}
where $\ee>0$ is a small parameter.
The Lagrangian density of the conventional scalar field theory can be written in
these coordinates as follows \cite{Ilgen, NPPF}:
\disn{4}{
L(y) = \dd_0\ff(y) \dd_1\ff(y) + \frac{\,\ee^2}{2}(\dd_0\ff(y))^2 - \frac{1}{2}(\dd_{\p}\ff(y))^2 -
\frac{ m_B^2}{2}(\ff(y))^2 - \la (\ff(y))^4,
\nom}
where $m_B$ is a mass parameter (the bare mass). The equation $y^0=0$ defines the space-like plane, so
the canonical
quantization on this plane is equivalent to the ordinary quantization on the $x^0=0$ plane in
Lorentz coordinates.
From the Lagrangian (\ref{4}) we obtain the following Hamiltonian density:
\disn{5}{
{\cal H} = \frac{(\Pi-\dd_1\ff)^2}{2\ee^2} + \frac{1}{2}(\dd_{\p}\ff)^2 + \frac{\,m_B^2}{2}\ff^2 + \la
\ff^4,
\nom}
where $\Pi(y)$ is the momentum canonically conjugated to the field $\ff(y)$, the $\Pi(y)=\ee^2 \dd_0 \ff(y)
+ \dd_1 \ff(y)$.

Further we consider the transition from the theories with the Hamiltonians
(\ref{5}) taken at different values of
the parameter $\ee$, to the LF Hamiltonian in the limit $\ee \to 0$. This gives
a possibility to take into account (before reaching the LF) two
different vacua existing in this model. Indeed,
at $\ee > 0$ we still can use known methods \cite{Weinberg} for the description
of the quantum vacuum. In particular we can apply the variational method \cite{Stev} to find
the minimum of the vacuum average of the Hamiltonian density. This method uses
different Fock vacua and Bogolyubov transformations from one Fock vacuum
to another (this method corresponds to the "Gaussian" variational approximation to the
vacuum wave function).
Let us apply this method to the Hamiltonian density (\ref{5}). We introduce the following expressions
for $\ff$ and $\Pi$ (at $y^0=0$):
\disn{6}{
\ff(y) = \frac{1}{2\pi} \int \frac{dk_1 dk_{\p}}{\sqrt{2\om(k)}} \Bigl( a(k) + a^+(-k) \Bigr)
e^{-ik\cdot y} + \ff_0,
\nom}
\disn{7}{
\Pi(y) = \frac{-i}{2\pi} \int dk_1 dk_{\p} \sqrt{\frac{\om(k)}{2}} \Bigl( a(k) - a^+(-k) \Bigr)
e^{-ik\cdot y},
\nom}
where $k=(k_1,k_{\p})$ and $k\cdot y=k_1y^1+k_{\p}y^{\p}$. Here we define the creation and annihilation
operators corresponding to the varying Fock vacua $|0\rangle$:
\disn{7a}{
a(k)|0\rangle = 0, \,\,\,
[a(k),a^+(k^{'})] = \delta^{(2)}(k-k^{'}), \,\,\, [a(k),a(k^{'})] = 0.
\nom}
The parameters $\om(k)$ and $\ff_0$ in (\ref{6}), (\ref{7}) play the role of variational
parameters (the $\ff_0$ doesn't depend on $k$). The variation of the parameters $\om(k)$ and $\ff_0$ is
equivalent to
linear transformations of operators $a, a^+$ that is equivalent to the variation of the vacuum state
vector $|0\rangle$ in the assumed approximation.
We implicitly suppose that the integration domain in the $k_1$ is limited by the cutoff $|k_1|\geqslant
\de$.
It is related to the necessity to get in the limit $\ee \to 0$ the theory on the LF which is regularized
by the cutoff $|k_-|\geqslant \de$.
Further we substitute the expressions (\ref{6}) and (\ref{7}) into the Hamiltonian (\ref{5}) and use the
equalities (\ref{7a}). We obtain the following result:
\disn{8}{
\langle0|{\cal H}|0\rangle = \frac{1}{16 \pi^2 \ee^2} \int dk_1 dk_{\p} \ls \om(k) + \frac{k_1^2 +
\ee^2(m_B^2 + k_{\p}^2 + 12\la\ff_0^2)}{\om(k)} \rs + \ns
+ \frac{\,m_B^2}{2}\ff_0^2 + \la \ff_0^4 + 3\la \ls \frac{1}{8\pi^2} \int \frac{dk_1 dk_{\p}}{\om(k)}
\rs^2.
\nom}
This expression contains divergent integrals. So we introduce the regularization of these integrals by
a cutoff in the momenta. Varying the quantity (\ref{8}) w.r.t. $\om(k)$ and equating the result to zero
we get
\disn{8a}{
\frac{1}{16 \pi^2 \ee^2} \ls 1 - \frac{k_1^2 + \ee^2 \ls m_B^2 + k_{\p}^2 + 12\la\ff_0^2 \rs}{\om^2(k)} -
\frac{3\la \ee^2}{2\pi^2 \om^2(k)} \int \frac{dq_1 dq_{\p}}{\om(q)} \rs = 0.
\nom}
Using the definition
\disn{8b}{
m^2 \equiv m_B^2 + 12\la\ff_0^2 + \frac{3\la}{2\pi^2} \int \frac{dq_1 dq_{\p}}{\om(q)},
\nom}
we obtain
\disn{9}{
\om^2(k) = k_1^2 + \ee^2 (m^2 + k_{\p}^2).
\nom}
Below we show that $m^2$ can be chosen to be finite in the regularization removing limit.

\begin{figure}[h!]
  \includegraphics[width=154mm]{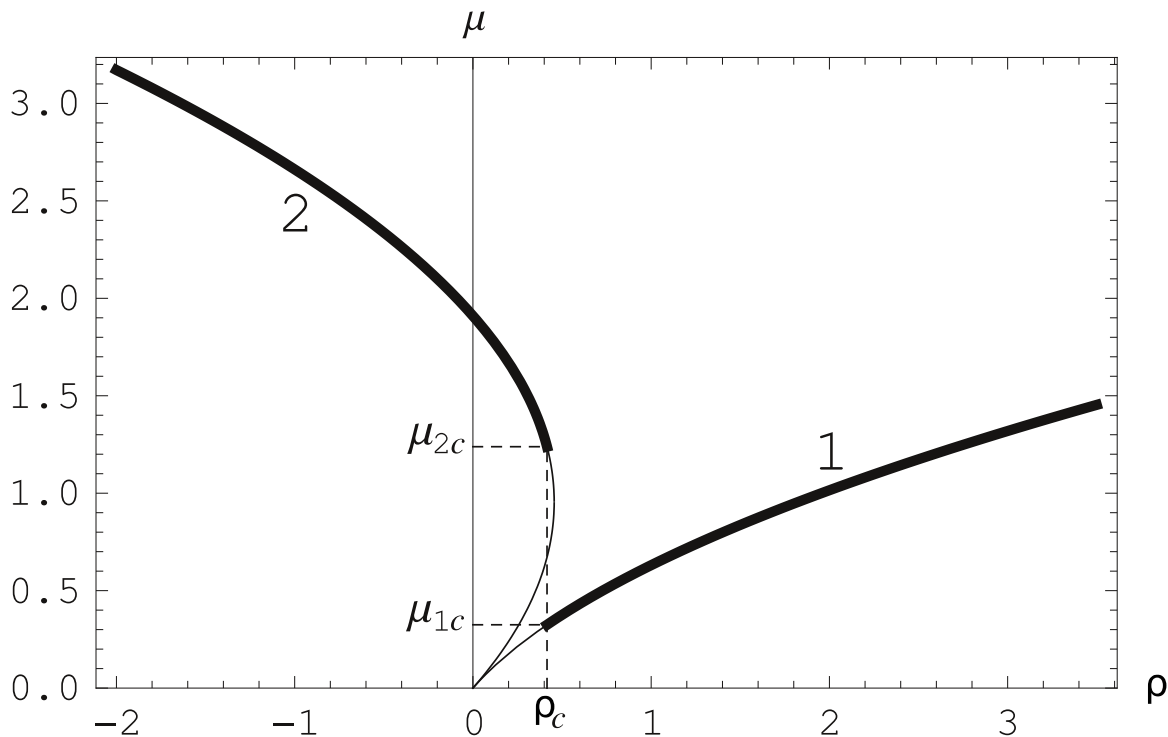}
  \caption{The dependence of $\m=\frac{m}{\la}$ on $\rho=\frac{r}{\la^2}$. Quantities $r$ and $m$
   are
defined by Eqs.~(\ref{9b}) and (\ref{10}). The curves 1 and 2
represent the solutions $\m_1(\rho)$ and $\m_2(\rho)$ of
Eq.~(\ref{14}). The bold curves show where these solutions
correspond to the minimum of vacuum energy density~(\ref{8}). The
$\rho_c$ is the point where this minimum is common for both
solutions. The $\m_{1c}$, $\m_{2c}$ are limit values of
$\m_1(\rho)$, $\m_2(\rho)$ as we approach to $\rho_c$ along
bold parts of curves.  We find  from expressions (\ref{14a}), (\ref{14b})
 $\rho_c\simeq0.4157$, $\m_{1c}\simeq0.3248$ and
$\m_{2c}\simeq~1.2385$.}
  \label{fig:1}
\end{figure}

The variation of Eq.~(\ref{8}) with respect to $\ff_0$ gives the equation
\disn{9a}{
\ff_0 \ls m_B^2 + 4\la\ff_0^2 + \frac{3\la}{2\pi^2} \int \frac{dk_1 dk_{\p}}{\om(k)} \rs = 0
\nom}
which can be rewritten in the following form (here we use the definition (\ref{8b})):
\disn{9aa}{
\ff_0 (m^2 - 8\la\ff_0^2) = 0.
\nom}
The solutions of this equation are $\ff_0 = 0$ and $\ff_0^2 = \frac{\,m^2}{8\la}$. One can check that
these solutions correspond to the minimum of the $\langle0|{\cal H}|0\rangle$ at $m^2>0$. Let us choose
the bare mass $m_B$ so that the parameter $m$ be finite:
\disn{9b}{
m_B^2 = - \frac{3\la}{2\pi^2} \int \frac{dk_1 dk_{\p}}{\sqrt{k_1^2 + \ee^2 k_{\p}^2}} + r,
\nom}
where the $r$ is finite in the regularization removing limit. Then the Eq.~(\ref{8b}) takes the
following form:
\disn{10}{
m^2 = 12\la\ff_0^2 + \frac{3\la}{2\pi^2} \int dk_1 dk_{\p}
\ls \frac{1}{\sqrt{k_1^2 + \ee^2 (k_{\p}^2 + m^2)}} - \frac{1}{\sqrt{k_1^2 + \ee^2 k_{\p}^2}}\rs + r.
\nom}
The integral in the Eq.~(\ref{10}) is convergent and needs no regularization. Then by the change of the
variable $k_1 \to \ee k_1$ one can reduce this integral
to a simpler form for which the result (in the $\de \to 0$ limit) is already known and equals to $-2\pi
m$. Let us define $\m = \frac{m}{\la}$ and $\rho = \frac{r}{\la^2}$. Then the Eq.~(\ref{10}) can be
rewritten in the following form
\disn{14}{
\m^2 + \frac{3\m}{\pi} - \frac{12\ff_0^2}{\la} - \rho = 0.
\nom}
We denote the solution of this equation for the case $\ff_0 = 0$ by $\m_1(\rho)$, and
for the case $\ff_0^2 = \frac{\,m^2}{8\la}$ by $\m_2(\rho)$. These solutions are shown in
Fig.~\ref{fig:1}.
The curves 1 and 2 show the solutions $\m_1(\rho)$ and $\m_2(\rho)$ correspondingly. We
consider these solutions at $\m>0$. For any $\rho$ in the domain
$0 < \rho \leqslant \frac{9}{2\pi^2}$ (the largest value corresponds to the rightmost point of
the curve 2) there are
several distinct values of $\m$ on the branches of curves with $\m>0$.
The direct evaluation of the quantity~(\ref{8}) shows that its minimum corresponds to points on the
 bold curves in Fig.~\ref{fig:1}.
Indeed, we consider the r.h.s. of Eq.~(\ref{8}) for the curve 1 and upper part of the curve 2 at
common value of $\rho$ and take the difference of these expressions.
Using Eq.~(\ref{8b}) and Eq.~(\ref{9b}) we find  the following finite result for this difference in
the regularization removing limit\footnote{At the first step we represent the expression for the
integral in Eq.~(\ref{8b}) through $m_B^2$, $m^2$, $\ff_0^2$, $\la$ and then use
this representation in Eq.~(\ref{8}).}:
\disn{14a}{
\frac{\la^3}{2}\ls\frac{1}{16}\ls2\m_1^4+\m_2^4\rs+\frac{1}{3\pi}
\ls\m_1^3-\m_2^3\rs+\frac{\rho}{12}\ls\m_2^2-\m_1^2\rs\rs.
\nom}
We estimate this expression numerically at different values of $\rho$ taking into account
the explicit dependence of $\m_1$, $\m_2$ on $\rho$ according to Eq.~(\ref{14}). We find that this expression is positive at $\rho<\rho_c$ where $\rho_c$ is the value of $\rho$ for which the expression (\ref{14a}) is equal zero. For $\rho_c<\rho\leqslant\frac{9}{2\pi^2}$ this expression is negative. The $\m_{1c}$ and $\m_{2c}$ are the limit values of $\m_1(\rho)$ and $\m_2(\rho)$ in the limit transition $\rho\to \rho_c$ along bold parts of curves 1 and 2 respectively. Numerically we find $\rho_c\simeq0.4157$, $\m_{1c}\simeq0.3248$ and $\m_{2c}\simeq~1.2385$.

Analogous comparison for corresponding lower and upper points on the curve~2 gives the following
expression:
\disn{14b}{
\frac{\la^3}{2}\ls\frac{1}{16}\ls\m_2^4-\bar{\m}_2^4\rs+\frac{1}{3\pi}
\ls\bar{\m}_2^3-\m_2^3\rs+\frac{\rho}{12}\ls\m_2^2-\bar{\m}_2^2\rs\rs,
\nom}
where $\bar{\m}_2$ denotes lower point on the curve 2. Analogously we find numerically that this expression is positive at
$\rho<\frac{9}{2\pi^2}$.

Thus we prove that the minimum of vacuum energy corresponds to the points on the bold curves. Therefore we have the
following inequalities limiting the parameters $\la$, $m_1\equiv\la\m_{1}$, $m_2\equiv\la\m_{2}$ which one should use in calculations with our Hamiltonian:
\disn{15}{
\frac{\la}{m_1} < \frac{1}{\m_{1c}}, \quad  \text{i.e.} \quad \m_1 > \m_{1c} \quad \text{for} \quad \ff_0 = 0;\no
\frac{\la}{m_2} < \frac{1}{\m_{2c}}, \quad \text{i.e.} \quad \m_2 > \m_{2c}  \quad \text{for} \quad
\ff_0^2 = \frac{\,m_2^2}{8\la}.
\nom}

Let us apply these results to the Hamiltonian (\ref{5}). We define the $\ti{\ff} = \ff - \ff_0$ and
rewrite the Hamiltonian in the normal ordered form w.r.t. those operators $a(k)$ and $a^+(k)$ which
correspond to the found vacuum.
Owing to Eq.~(\ref{8b}) the resulting expression becomes dependent on the mass parameters $m_1$, $m_2$
only. These parameters correspond to solutions shown in Fig.~\ref{fig:1}.
In the case $\ff_0=0$ we get the following Hamiltonian (throwing out the constant term $\langle0|{\cal
H}|0\rangle$):
\disn{16}{
H = \, :\int dy^1 dy^{\p} \ls \frac{(\Pi-\dd_1\ti{\ff})^2}{2\ee^2} +
\frac{1}{2}\,(\dd_{\p}\ti{\ff})^2 + \frac{\,m_1^2}{2}\ti{\ff}^2 + \la \ti{\ff}^4 \rs:,
\nom}
where the symbol ": :" denotes the normal ordering. Analogously, in the case $\ff_0^2 =
\frac{\,m_2^2}{8\la}$ we obtain
\disn{17}{
H = \, :\int dy^1 dy^{\p} \ls \frac{(\Pi-\dd_1\ti{\ff})^2}{2\ee^2} + \frac{1}{2}\,(\dd_{\p}\ti{\ff})^2
 + \frac{\,m_2^2}{2}\ti{\ff}^2 + 4 \la \ff_0 \ti{\ff}^3 + \la \ti{\ff}^4 \rs:.
\nom}
Here the terms linear in the fields $\ti{\ff}, \Pi$ are discarded because they don't contribute to the
integral (\ref{17}) due to
the condition $|k_1|\geqslant\de>0$ proposed earlier for the integration in formulae (\ref{6}) and
(\ref{7}) (see the text before Eq.~(\ref{8})).

To find the form of the LF Hamiltonian let us consider the eigenvalue problem:
\disn{18}{
H |f\rangle = E |f\rangle,
\nom}
where the $H$ is the Hamiltonian (\ref{16}) or (\ref{17}). One can expand these Hamiltonians in powers
of the parameter $\ee$.
We separate the $\ee^{-2}$ term of these Hamiltonians and write them in the form
\disn{19b}{
H = \frac{H_0}{\ee^2} + H_2,
\nom}
where
\disn{19}{
H_0 = 2 \int_{-\infty}^0 dk_1 \int dk_{\p} |k_1| \, a^+(k) \, a(k).
\nom}
In the derivation of this expression we use the equality $\om(k)=|k_1|+\frac{\ee^2(m^2
+k_{\p}^2)}{|2k_1|}+~O(\ee^4)$
following from the Eq.~(\ref{9}). Let us write the following asymptotic expansions:
\disn{19a}{
E(\ee)  = \frac{E_0}{\ee^2} + E_2 + \cdots, \qquad |f(\ee)\rangle = |f_0\rangle + \ee^2 |f_2\rangle +
\cdots
\nom}
In the lowest order approximation w.r.t. $\ee$ we obtain the equations:
\disn{20}{
H_0 |f_0\rangle = E_0 |f_0\rangle \, , \qquad (H_0 - E_0) |f_2\rangle + (H_2 - E_2) |f_0\rangle = 0.
\nom}
In the limit $\ee \to 0$ we have $x^1 \to x^-, \,\, |f\rangle \to |f_0\rangle$, i.e. the states
$|f_0\rangle$ form the state space on the LF.
To get finite eigenvalues for the LF Hamiltonian we demand $E_0 = 0$. Then from the Eq.~(\ref{19}) and
the first of Eqs.~(\ref{20}) we obtain
\disn{21}{
a(k)|f_0\rangle = 0 \quad \text{at} \quad k_1<0.
\nom}
Therefore in the limit $\ee \to 0$ the LF state space is the subspace of our Fock space in which only
the quanta with $k_->0$ are present. Now let us take the projection of the second of the Eqs.~(\ref{20})
on the subspace of states $|f_0\rangle$ and denote by ${\cal P}$ the projector on this subspace. Then we
get the equation which can be interpreted as the eigenvalue equation for the LF Hamiltonian. So now we
have
\disn{23}{
H_{LF} = {\cal P} H_2 {\cal P}.
\nom}
Using the expressions (\ref{16}) and (\ref{17}) and the equality (\ref{21}) we obtain the following
results:
\disn{24}{
H_{LF} = \, :\int dx^- dx^{\p} \ls \frac{1}{2}\,(\dd_{\p}\Phi)^2 + \frac{\,m_1^2}{2}\Phi^2 + \la \Phi^4
\rs:
\nom}
for the case $\ff_0 = 0$ and
\disn{26}{
H_{LF} = \, :\int dx^- dx^{\p} \ls \frac{1}{2}\,(\dd_{\p}\Phi)^2 + \frac{\,m_2^2}{2}\Phi^2 + 4 \la \ff_0
\Phi^3 + \la \Phi^4 \rs:
\nom}
for the case $\ff_0^2 = \frac{\,m_2^2}{8\la}$. Here we denote by $\Phi(x)$ the field on the LF,
\disn{25}{
\Phi(x) = \frac{1}{2\pi} \int_{\de}^{\infty} \frac{dk_-}{\sqrt{2k_-}} \int dk_{\p}
\ls a(k_-,k_{\p}) \, e^{-ik\cdot x} + a^+(k_-,k_{\p}) \, e^{ik\cdot x} \rs,
\nom}
where $k\cdot x = k_-x^-+k_{\p}x^{\p}$. The operators $a^+(k_-,k_{\p})$ and $a(k_-,k_{\p})$ play the
role of creation and annihilation operators in the LF Fock space. They satisfy canonical commutation
relations on the LF. Note that the integration range in Eq.~(\ref{25}) is limited from below
by a small parameter $\de$ which we implicitly use in the Eqs.~(\ref{6}), (\ref{7}) (see the text before
Eq.~(\ref{8})).

\section{Investigation of perturbation theory}
\label{sec:2}
\begin{figure}[h!]
  \includegraphics[width=\textwidth]{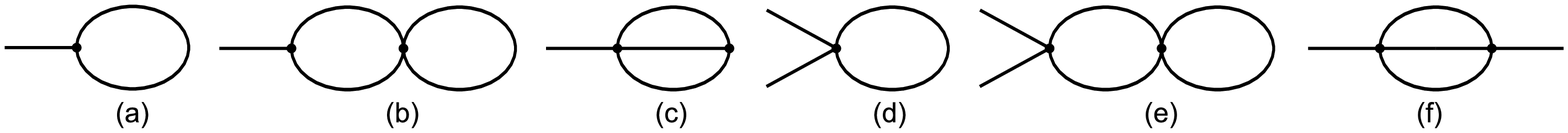}
  \caption{The divergent diagrams.}
\label{fig:2}
\end{figure}

In the previous section we have obtained in Gaussian approximation the LF Hamiltonians (\ref{24}),
(\ref{26}).
The theories described by these LF Hamiltonians contain UV divergences. To study these divergences and
carry out the renormalization let us compare the perturbation theories, generated by these LF
Hamiltonians, with the corresponding renormalized covariant perturbation theories in Lorentz coordinates
(in all orders). With this aim we consider the covariant perturbation theory for the Lagrangian in the
following general form (in Lorentz coordinates):
\disn{s1}{
{\cal L}=\frac{1}{2}\dd_\m\ff \dd^\m\ff-\frac{m^2}{2}\ff^2-\g\ff-g\ff^3-\la\ff^4.
\nom}
Standard analysis shows that this theory is superrenormalizable, i.e. it has only finite number of
divergent diagrams
which must be renormalized. These diagrams are shown in Fig.~\ref{fig:2}.
The sum of all connected diagrams with one external line ("tadpole" diagrams) gives the one-point Green
function. It is a constant
in the coordinate space and is equal to the vacuum average of the field $\ff$. So one can make
this constant to be equal to zero by the shift of the field: $\ff\to\ff-\ff_0$, where $\ff_0$ is the
constant. In this way we obtain the theory without tadpole diagrams (and subdiagrams).
The linear in the field term of the Lagrangian (\ref{s1}) generates only the tadpole subdiagrams.
Therefore if we consider the perturbation theory without tadpole subdiagrams, we can ignore this term in
the Lagrangian.
The parameter $\la$ in the Lagrangian (\ref{s1}) does not change after the shift in the field, while the
parameters $m^2$ and $g$ do change.

The diagrams with two external lines joined to one vertex (we denote them by \mbox{"tadpole-2"{}}
diagrams) do not depend on external momenta. All such connected diagrams are one-particle-irreducible
ones (in the absence of the tadpole subdiagrams), and their contribution is equivalent to the
addition of the counterterm, quadratic in the field, to the Lagrangian. In this way one can formulate
the perturbation theory in Lorentz coordinates without
tadpole and \mbox{tadpole-2} diagrams (and corresponding subdiagrams).

Then we have only one logarithmically divergent diagram, the divergent part of which must be compensated
by the counterterm quadratic in the field.
This diagram is shown in Fig.~\ref{fig:2}~(f), and in the following we denote it by $I(p)$. Further we
choose the PV method for the regularization of the obtained perturbation theory. This choice requires the
introduction of the auxiliary large mass $M$ ghost field $\ff_g(x)$ into the Lagrangian.

Thus we can generate the considered perturbation theory in Lorentz coordinates by the
following regularized Lagrangian if we throw away all tadpole and \mbox{tadpole-2} diagrams:
\disn{s2}{
{\cal L}=\frac{1}{2}\ls\dd_\m\ff \dd^\m\ff-\tilde m^2\ff^2\rs-\frac{1}{2}\ls\dd_\m\ff_g
\dd^\m\ff_g-M^2\ff_g^2\rs-
\frac{3\la^2}{\pi^2} \ls \ln \frac{M}{\tilde m} \rs \ls\ff+\ff_g\rs^2-\ns
-\tilde g\ls\ff+\ff_g\rs^3-\la\ls\ff+\ff_g\rs^4.
\nom}
Here the mass $M$ is the regularization parameter, and the quadratic in the field $(\ff+\ff_g)$
counterterm
is added to compensate the divergent part of the diagram $I(p)$ at $M\to\infty$. This divergent part is
calculated in Appendix A.

Now let us write the canonical Hamiltonian on the LF, corresponding to this Lagrangian,
and take this Hamiltonian in the normally ordered form (in accordance with the absence of the tadpole
and \mbox{tadpole-2} diagrams
in the previously considered perturbation theory in Lorentz coordinates):
\disn{s3}{
H_{LF}=\,:\int dx^-dx^\p\Biggl(
\frac{1}{2}\ls\dd_{\p}\Phi\rs^2-\frac{1}{2}\ls\dd_{\p}\Phi_g\rs^2
+ \frac{\,\tilde m^2}{2}\,\Phi^2 - \frac{\,M^2}{2}\,\Phi_g^2+\ns
+\frac{3\la^2}{\pi^2} \ls \ln \frac{M}{\tilde m} \rs\ls\Phi+\Phi_g\rs^2
+\tilde g\ls\Phi+\Phi_g\rs^3+\la\ls\Phi+\Phi_g\rs^4\Biggr):.
\nom}
Here we introduce, like in the Eq.~(\ref{25}), the regularization parameter $\de$ for the Fourier
decomposition of fields $\ff$ and $\ff_g$ in terms of creation and annihilation operators on the LF,
and again denote these regularized fields by $\Phi$ and $\Phi_g$. The canonical commutation relations
for the ghost creation and annihilation operators
have the opposite sign w.r.t. conventional ones\footnote{One can find out that the interaction terms in
this LF Hamiltonian remain normal ordered even
if one removes normal ordering symbol ": :" in the Eq.~(\ref{s3}).}.

Starting from this Hamiltonian we can generate the same set of diagrams as in covariant perturbation
theory in Lorentz coordinates.
Indeed, it was proven in paper~\cite{hep-ph/9412315} that
each term in covariant perturbation theory (i.e. Feynman diagram) can be written as the sum of
terms
in the $x^+$-ordered ("old-fashioned") perturbation theory series.
But the difference is contained in the way of calculation of diagrams:
for LF perturbation theory the integration must be carried out firstly over the momenta $k_+$ and
the condition $|k_-|\geqslant\de$ must be introduced as the regularization of fields in
Eq.~(\ref{s3}).

Let us remark that this way of calculation gives all tadpole and \mbox{tadpole-2} diagrams equal to zero.
Indeed, the tadpole diagrams are absent because the momentum of the external line of these diagrams is
equal to zero
but the regularization of fields on the LF ($|k_-|\geqslant\de$) doesn't allow such external momentum.
\mbox{Tadpole-2} diagrams are equal zero because in such a diagram there is always a closed loop that
has the same sign of the loop momentum $q_-$ in all its propagators. As the result the residue integral
w.r.t. corresponding loop momentum $q_+$
is equal to zero, because all poles related to these propagators lie on the same side of the real axis
of $q_+$.
Two divergent \mbox{tadpole-2} diagrams shown in Fig.~\ref{fig:2}~(d),~(e) are absent due to normal
ordering of the Hamiltonian (\ref{s3}).
So the set of nonzero diagrams turns out to be the same in perturbation theory in Lorentz coordinates
and that generated by the Hamiltonian on the LF.

However it is known that the corresponding diagrams calculated in each of these perturbation theories
can differ \cite{burlang, tmf97, NPPF}. One can use the method of the papers \cite{tmf97, NPPF} to
compare such diagrams in all orders of perturbation theory.
The idea of this method is the following. The diagrams generated by the LF Hamiltonian are regularized
with the $|k_-|\geqslant\de$ cutoff.
Therefore the difference between the result of calculation of such a diagram and the result of
calculation of
the corresponding covariant diagram in Lorentz coordinates reduces to the integrals over the region
$|k_-|\leqslant\de$ of each propagator. If one makes for each loop momentum $q$ (which can be always
identified with a some propagator momentum) the change $q_-\to q_- \de$, $q_+\to q_+/\de$ an essential
dependence on $\de$ in the integration region disappears, and one can investigate the behavior of the
integrand at $\de\to 0$ for an arbitrarily complicated diagram using only
general properties of the theory (Lorentz invariance, the spin of the field, the structure of the
propagator, etc.). With this method it is possible to prove for our model that
the results of calculation
of any diagram in the LF perturbation theory and in the covariant perturbation theory in Lorentz
coordinates coincide in the limit $\de\to0$ (taking into account the absence of tadpole and
\mbox{tadpole-2} diagrams).
In Appendix B we illustrate how this method works with the example of one-loop diagram.

Thus the theory with the LF Hamiltonian (\ref{s3}) turns out to be equivalent in all orders
of perturbation theory to the conventional renormalized covariant perturbation theory in Lorentz
coordinates in the limit $\de\to 0$ (and then $M\to\infty$).

Therefore the Eq.~(\ref{s3}) gives the
perturbatively renormalized LF Hamiltonian. We notice that the LF Hamiltonian (\ref{s3}) can be
considered, at definite choice of its parameters, as one of the LF Hamiltonians (\ref{24}),
(\ref{26}), correspondingly regularized and renormalized. The coupling constant $\tilde g$ can be
identified with $4\la\ff_0$ ($\ff_0=0$ or $\ff_0^2=\frac{m_2^2}{8\la}$), while $\tilde m$ can be
identified with $m_1$ or $m_2$ for the LF Hamiltonians (\ref{24}) and (\ref{26}) respectively.
Thus we obtain the renormalized LF Hamiltonians for the cases without and with the spontaneous
symmetry breaking. The parameters $m_1$, $m_2$ and $\la$ satisfy the inequalities (\ref{15}) in
the Gaussian approximation. Nevertheless one can assume that these inequalities are approximately
valid for the renormalized LF Hamiltonians in the PV regularization.

\section{Conclusion}
\label{concl}
In the present paper we have constructed the renormalized LF Hamiltonian for the $\la \ff^4$ model
in (2+1)-dimensional space-time.
We have found the explicit expression for the counterterm, necessary for the renormalization,
using the PV regularization. To do this we compare the diagrams of the covariant perturbation
theory in Lorentz coordinates with the analogous diagrams of the perturbation
theory generated by the LF Hamiltonian which has also the cutoff in the momentum
$ p_-$ ($|p_-|\geqslant \delta >0$). We show that both perturbation theories can be described by
the same set of diagrams, with the values of the compared diagrams coinciding in the limit
$\delta \to 0 $. Then we renormalize the LF Hamiltonian by the counterterm found in
the calculation of the divergent part of the corresponding diagram in the covariant perturbation
theory in Lorentz coordinates.

Furthermore we have  taken into account the possibility of the spontaneous symmetry
breaking in this model and obtained the LF Hamiltonians corresponding to two different
vacua. We arrive at these LF Hamiltonians by considering the limit transition from the
theories quantized on the spacelike planes approaching the LF. It is
possible to describe the vacuum on these planes using the Gaussian approximation.
The Hamiltonians obtained with this approximation still require UV renormalization.
And the above-mentioned comparison of perturbative theories, generated by these
LF Hamiltonians, and the covariant perturbation theory in Lorentz coordinates allows
to renormalize both of these Hamiltonians in the PV regularization.

Having such LF Hamiltonians one can start nonperturbative calculation
of the mass spectrum solving the eigenvalue problem:
\disn{28}{
(2P_-H_{LF} - P_{\bot}^2)|p_-,p_{\bot}\rangle = m^2 |p_-,p_{\bot}\rangle.
\nom}
This calculation could help to study the peculiarities of the PV
regularization related to the presence of ghost fields and states with the
indefinite metric. That is important for the application of this method
to QCD.

To solve the eigenvalue problem (\ref{28}) nonperturbatively
one can apply numerical approach using the discretization of the momenta:
$p_- = \frac{\pi n}{L}, \, n = 0,1,2,\ldots$; $p_{\p} = \frac{\pi n_{\p}}{L_{\p}}$, ${n_{\p}=0, \pm1,
\pm2, \ldots}$
This discretization is achieved by introducing the
limits ${|x^-|\leqslant L}$, $|x^{\p}|\leqslant L_{\p}$ and corresponding
periodic boundary conditions for fields (see e.g. \cite{Brodsky1, Brodsky3, Brodsky4}).
Also it is useful to introduce the lattice in $x^{\p}$ to limit the $n_{\p}$.
This calculation must be carried out at finite but increasing values of the
parameters $M, L, L_{\p}$ and the cutoff in $n_{\p}$.
As it was shown in the papers \cite{tmf97, NPPF}
the regularization $|p_-|\geqslant \delta >0$ must be removed before the
removing of PV regularization. In the DLCQ (with zero modes ($p_-=0$) being thrown out) one can put
$\delta=\frac{\pi}{L}$. So we must at first take $L\to\infty$. For finite
total momentum $p_- = \frac{\pi n}{L}$
this limit is equivalent to $n\to\infty$.
The successful calculation of few lowest values of mass in QED(1+1) \cite{Yad.Fiz.2005}
for all values of the coupling (including very large ones)  shows the possibility to make the
extrapolation
to $n\to\infty$ using the results obtained at finite values of $n$.
One can assume that such extrapolation may be done at every fixed values of $M$, $L_{\p}$
and lattice parameter.
It must be noticed that in such calculations zero modes of fields ($p_-=0$), playing important
role for correct description of possible vacuum effects \cite{Hornbostel.Phys.Rev.D1992}, are
excluded because in the
considered scalar field theory the effect of spontaneous symmetry breaking can be approximately
 taken into account.
 In gauge theories like QCD these zero modes
can be included into LF canonical formalism
\cite{Franke.Novozhilov.Prokhvatilov.Lett.Math.Phys.1981} where they are to be defined
via solution
 of complicated second class constraints that remains the difficult problem.
   Correct description of vacuum effects with the LF Hamiltonian can be given  for
  QED(1+1) \cite{tmf02, Yad.Fiz.2005} and the role of zero modes can be seen there.

The generalization of these methods to gauge theories requires the construction of renormalized LF Hamiltonians for these theories.
A possible way to construct such Hamiltonians in PV regularization was proposed for light-cone gauge QCD(3+1) in \cite{tmf99, NPPF} and applied for calculation of anomalous magnetic moment in QED(3+1) \cite{Brodsky.Franke.Hiller.McCartor.Paston.Prokhvatilov2004}.

The AdS/QCD correspondence and its relation to QCD bound state problem
\cite{Brodsky.Teramond.Phys.Rev.Lett.2006, Brodsky.Teramond.Phys.Rev.Lett.2009,
Brodsky.Teramond.Dosch.Phys.Rev.D2013, Brodsky.Teramond.Dosch.arXiv2013}
suggests the description of bound state wave functions in terms of some special
functions different from usual plane wave ones.
This implies that the basis functions used in the decomposition of fields
in terms of light-front creation and annihilation operators can be chosen
in accordance with those special functions in a hope to make the solution
of Eq.~(\ref{28}) more effective.
This approach was applied in papers \cite{Vary.Brodsky.Harindranath.Teramond.et.al.arXiv2009,
Li.Wiecki.Zhao.Maris.Vary.arXiv2013,
Vary.Zhao.Ilderton.Honkanen.Maris.Brodsky.Acta.Physica.Polonica.B2013}.

Despite of difficulties these methods meet one hopes that the LF quantization could give frame-invariant
and unified scheme for the description
of hadron physics at high energies and, nonperturbatively, at lower
energies \cite{Bakker.Bassetto.Brodsky.Broniowski.Dalley.Frederico.Glazek.
Hiller.Ji.Karmanov.et.al.arXiv2013}.

\vskip 0.5em
{\bf Acknowledgments.}
We thank M.V.~Kompaniets and V.A.~Franke for useful discussions.
The authors M.Yu.M., E.V.P. and R.A.Z. acknowledge Saint-Petersburg State University for a research grant 11.38.189.2014.

\section*{Appendix A. The calculation of the diagram $I(p)$}
\label{app:A}
For the renormalization of our model in the conventional Feynman covariant perturbation
theory it is necessary to consider only one logarithmically divergent diagram (Fig.~\ref{fig:2}~(f)).
This diagram in the PV regularization has the following form:
\disn{32}{
I(\vec{p}) = \frac{96i\la^2}{(2\pi)^6} \int d^3\vec{k}_1 \, d^3\vec{k}_2 \, d^3\vec{k}_3 \, \de \ls
\sum_{j=1}^3 \vec{k}_j - \vec{p} \rs \prod_{j=1}^3
\ls \frac{1}{\vec{k}_j^2 + m^2} - \frac{1}{\vec{k}_j^2 + M^2} \rs,
\nom}
where the integration is over the Euclidean momenta, the parameter $m$ is the mass parameter, the $M$ is
PV regularization parameter, the factor $96=(4!)^2/6$ includes the symmetry coefficient 1/6 of this
diagram (the factor $(4!)^2$ is related to the definition of the coupling $\la$ in the Lagrangian
(\ref{s1})).
To find the counterterm we need to calculate only the divergent (at $M \to\infty$) part of the
$I(\vec{p})$.
This divergent part can be evaluated as the divergent part of the $I(\vec{p})|_{\vec{p}=0}$ which can be
rewritten in the following form:
\disn{a35}{
I(0) = \frac{96i\la^2}{(2\pi)^9} \int d^3\vec{x} \,\, e^{i\vec{x}(\vec{k}_1+\vec{k}_2+\vec{k}_3)}
\prod_{j=1}^3 \int d^3\vec{k}_j \ls \frac{1}{\vec{k}_j^2 + m^2} - \frac{1}{\vec{k}_j^2 + M^2} \rs = \ns
= \frac{3i\la^2}{16\pi^9} \int d^3\vec{x} \ls \int d^3\vec{k} \,\, e^{i\,\vec{k}\vec{x}}
\ls \frac{1}{\vec{k}^2 + m^2} - \frac{1}{\vec{k}^2 + M^2} \rs \rs^3.
\nom}
We can use the well-known result:
\disn{a36}{
\int d^3\vec{k} \frac{e^{i\vec{k}\vec{x}}}{\vec{k}^2 + m^2} = \frac{2\pi^2}{r} e^{-rm}, \quad
r=\sqrt{\vec{x}^2}.
\nom}
To obtain this result it is sufficient to put $\vec{x}=\{r,0,0\}$ and evaluate residue integral:
\disn{a37}{
\int d^3\vec{k} \frac{e^{ik_1r}}{k_1^2+ k_2^2+ k_3^2 + m^2} = 2\pi i \int dk_2\,dk_3
\frac{e^{-r\sqrt{k_2^2+ k_3^2 + m^2}}}{2i\sqrt{k_2^2+ k_3^2 + m^2}} = \frac{2\pi^2}{r} e^{-rm}.
\nom}
Further we substitute this result into Eq.~(\ref{a35}):
\disn{a35a}{
I(0)=\frac{6i\la^2}{\pi^2}\int_0^{\infty}\frac{dr}{r}\ls
e^{-rm}-e^{-rM}\rs^3=\frac{6i\la^2}{\pi^2}\biggl(\int_0^{1}\frac{dr}{r}\ls
e^{-r\frac{m}{M}}-e^{-r}\rs^3+\ns
+ \int_1^{\infty} \frac{dr}{r} e^{-3r\frac{m}{M}} + \int_1^{\infty} \frac{dr}{r} \ls
-3e^{-2r\frac{m}{M}}e^{-r}
+ 3e^{-r\frac{m}{M}}e^{-2r} - e^{-3r} \rs \biggr).
\nom}
The first and third integrals in this expression are finite constants at $M\to\infty$, and therefore,
the $I(0)$ can be rewritten as follows:
\disn{36}{
I(0) = \frac{6i\la^2}{\pi^2} \int_1^{\infty} \frac{dr}{r} e^{-3r\frac{m}{M}} + O(1) =
\frac{6i\la^2}{\pi^2} \int_{\frac{m}{M}}^{\infty} \frac{dr}{r} e^{-3r} + O(1) = \ns
= \frac{6i\la^2}{\pi^2} \ls \int_{1}^{\infty} \frac{dr}{r} e^{-3r}
+ \int_{\frac{m}{M}}^{1} \frac{dr}{r} \ls e^{-3r} - 1 \rs + \int_{\frac{m}{M}}^{1} \frac{dr}{r} \rs +
O(1).
\nom}
The first two integrals in this expression are finite constants at $M\to\infty$, thus we obtain for
$I(0)$ the following result:
\disn{37}{
I(0) = \frac{6i\la^2}{\pi^2} \ln{\frac{M}{m}} + O(1).
\nom}
Using this result one can find the corresponding counterterm in the standard way \cite{Weinberg}.
This counterterm is present in the Eqs.~(\ref{s2}) and (\ref{s3}).

\section*{Appendix B. The example of comparison of diagram calculations}
\label{app:B}
Let us demonstrate how the method of comparison of calculations in LF and usual covariant
perturbation theory \cite{NPPF, tmf97} works using as an example
the 1-loop diagram with two external lines.
Let us write the corresponding integrand in the following form:
\disn{B1}{
\frac{1}{\ls q^2-m^2+i0 \rs \ls (q+p)^2-m^2+i0 \rs}=\ns=
\frac{1}{\ls 2q_+q_--q_{\p}^2-m^2+i0 \rs\ls 2(q_++p_+)(q_-+p_-)-(q_{\p}+p_{\p})^2-m^2+i0 \rs},
\nom}
where $p$ and $q$ are external and loop momenta respectively.

In the calculation of diagrams in usual covariant perturbation theory the integration is carried out
over all momenta $q_{\m}$ while in the
LF calculation the integration is only over the domain
$\{|q_-|\geqslant\de\}\cap \{|q_-+p_-|\geqslant\de\}$ due to the regularization
of fields in Eq.~(\ref{s3}). So the difference between the LF and the conventional covariant
calculations of the diagram is given by the integral over the domain
$\{|q_-|<\de\}\cup\{|q_-+p_-|<\de\}$. This domain consists of two parts and the contribution of
each of them should be considered separately. However the  contribution of the second part becomes
similar to one of the first part after the change $q \to \tilde{q} = q + p $. So we consider
the integration only over the first part of the integration domain:
\disn{B2}{
\int dq_{\p}\int dq_+\int_{-\de}^{\de} dq_- \frac{1}{\ls 2q_+q_--q_{\p}^2-m^2+i0
\rs}\times\ns\times\frac{1}{\ls 2(q_++p_+)(q_-+p_-)-(q_{\p}+p_{\p})^2-m^2+i0\rs}.
\nom}
After the change $q_- \to q_-\de$, $q_+ \to q_+/\de$ this integral takes the form:
\disn{B3}{
\int dq_{\p}\int dq_+\int_{-1}^{1} dq_-\frac{1}{\ls 2q_+q_--q_{\p}^2-m^2+i0 \rs}\times\ns\times
\frac{\de}{\ls 2(q_++p_+\de)(p_-+q_-\de)-(q_{\p}+p_{\p})^2\de-m^2\de+i0\rs}.
\nom}
The integration domain is now independent on $\de$ while the expression for the integrand can be
easily analyzed in the $\de \to 0$ limit. We see that in this limit the integral (\ref{B3}) is equal
to zero.

If we add in the expression (\ref{B1}) the factor $q_+$ in the numerator
(as e.g. in fermion self-energy diagram in Yukawa model, see \cite{NPPF, tmf97}) we get after
the change $q_+ \to q_+/\de$ the additional factor $1/\de$. So we obtain nonzero result for
the difference between LF and conventional covariant calculations of such diagram.
We see that the dependence on the external momenta factorizes in this result for the considered
difference.
Such factorization usually has place also in higher orders of perturbation theory
\cite{NPPF, tmf97}.
This gives a possibility to determine the form of necessary counterterms which must be added to
the canonical LF Hamiltonian to remove the above-mentioned difference.

\end{document}